\documentstyle[prd,aps]{revtex}

\newcommand{\prepr}[1] {\begin{flushright}  {\bf #1} \end{flushright}
\vskip 1.cm}
\begin{document}
%
\prepr{
\hspace{20mm} NCKU-HEP-00-04 \\
\hspace{20mm} hep-ph/0006xxx }
\centerline{\large{\bf Final state interaction and $B\to KK$ decays in
perturbative QCD}}
\vskip 1.0cm
\centerline{Chuan-Hung Chen and Hsiang-nan Li}
\vskip 0.5cm
\centerline{Department of Physics, National Cheng-Kung University,}
\centerline{Tainan, Taiwan 701, Republic of China}


\vskip 1.0cm
\vspace{10mm}
\centerline{\bf abstract}
\vskip 0.5cm
We predict branching ratios and CP asymmetries of the $B\to KK$ decays
using perturbative QCD factorization theorem, in which tree, penguin, and
annihilation contributions, including both factorizable and
nonfactorizable ones, are expressed as convolutions of hard six-quark
amplitudes with universal meson wave functions. The unitarity angle
$\phi_3= 90^o$ and the $B$ and $K$ meson wave functions extracted from
experimental data of the $B\to K\pi$ and $\pi\pi$ decays are employed.
Since the $B\to KK$ decays are sensitive to final-state-interaction
effects, the comparision of our predictions with future data can test the
neglect of these effects in the above formalism. The CP asymmetry in
the $B^\pm\to K^\pm K^0$ modes and the $B_d^0\to K^\pm K^\mp$ branching
ratios depend on annihilation and nonfactorizable amplitudes. The
$B\to KK$ data can also verify the evaluation of these contributions.


\vskip 1.0cm

\section{INTRODUCTION}

The conventional approach to exclusive nonleptonic $B$ meson decays
relies on the factorization assumption (FA) \cite{BSW}, under which
nonfactorizable and annihilation contributions are neglected and
final-state-interaction (FSI) effects are assumed to be absent.
Factorizable contributions are expressed as products of Wilson
coefficients, meson decay constants, and hadronic transition form factors.
Though analyses are simpler under this assumption, estimations of
many important ingredients, such as tree and penguin (including
electroweak penguin) contributions, and strong phases are not reliable.
Moreover, the above naive FA suffers the problems of scale,
infrared-cutoff and gauge dependences \cite{CLY}. It is also difficult to
explain the observed branching ratios of the $B\to J/\psi K^{(*)}$ decays
in the FA approach, to which nonfactorizable and factorizable
contributions are of the same order \cite{YL}.

Perturbative QCD (PQCD) factorization theorem for exclusive heavy-meson
decays has been developed some time ago \cite{LY1,L1,CL}, which goes
beyond FA. PQCD is a method to separate hard components from a QCD process,
which are treated by perturbation theory. Nonperturbative components are
organized in the form of hadron wave functions, which can be extracted
from experimental data. This prescription removes the infrared-cutoff
dependence in PQCD. Since nonperturbative dynamics has been absorbed into
wave functions, external quarks involved in hard amplitudes are on-shell,
and gauge invariance of PQCD predictions is guaranteed. Contributions to
hard parts from various topologies, such as tree, penguin and
annihilation, including both factorizable and nonfactorizable
contributions, can all be calculated. Without assuming FA, it is easy to
achieve the scale independence in the PQCD approach.

Despite of the above merrits of PQCD, an important subject, final state
interaction (FSI), remains unsettled, which is nonperturbative but not
universal. FSI effects in two-body decays have been assumed to be small.
Though arguments and indications for this assumption have been supplied
in \cite{KLS}, experimental justification is necessary. In this
paper we shall propose to explore FSI effects by studying $B\to KK$
decays. It will be explained in Sec.~II that these decays are more
sensitive to FSI effects compared to $B\to K\pi$ and $\pi\pi$ decays.
Employing the meson wave functions and the unitarity angle
$\phi_3=90^o$ determined in \cite{KLS}, we predict the branching ratios
and the CP asymmetries of the $B^\pm\to K^\pm K^0$, $B_d^0\to K^\pm K^\mp$
and $B_d^0\to K^0 {\bar K}^0$ modes. The comparision of our predictions
with future data can be used to estimate the importance of FSI effects.
In particular, large observed $B_d^0\to K^\pm K^\mp$ branching ratios and
CP asymmetry in the $B_d^0\to K^0 {\bar K}^0$ modes will imply strong FSI
effects.

An essential difference between the FA and PQCD approaches is that
annihilation and nonfactorizable amplitudes are neglected in the former,
but calculable in the latter. It has been shown that annihilation
contributions from the operator $O_{5,6}$ with the $(V-A)(V+A)$ structure,
bypassing helicity suppression, are not negligible \cite{KLS}. These
contributions, being mainly imaginary, result in CP asymmetries in the
$B\to \pi\pi$ decays, which are much larger than those predicted in FA
\cite{LUY,KL}. Hence, measurements of CP asymmetries will distinguish the
two approaches \cite{KL}. The $B^\pm\to K^\pm K^0$ modes contain both
annihilation amplitudes from $O_{5,6}$ and nonfactorizable annihilation
amplitudes from $O_{1,2}$, such that they exhibit substantial CP asymmetry.
The branching ratios of the $B_d^0\to K^\pm K^\mp$ modes, involving only
nonfactorizable annihilation amplitudes, can not be estimated, or are
vanishingly small in FA. The data of these two decays can verify PQCD
evaluation of annihilation and nonfactorizable contributions.

FSI effects in the $B\to \pi\pi$, $K\pi$ and $KK$ decays are compared
in Sec.~II. The PQCD formalism for annihilation and nonfactorizable
contributions is reviewed in Sec.~III. We present the factorization
formulas of all the $B\to KK$ modes in Sec.~IV, and perform a numerical
analysis in Sec.~V. Section VI is the conclusion.

\section{FINAL STATE INTERACTION}

FSI is a subtle and complicated subject. Most estimates of FSI effects in
the literature \cite{FSI} suffer ambiguities or difficulties. Kamal has
pointed out that the enhancement of CP asymmetry in the
$B^{\pm} \to K^{0}\pi^{\pm}$ modes from order 0.5 \% up to order
(10-20)\% \cite{KPS} is due to an overestimation of FSI effects by a
factor of 20 \cite{Kamal}. The smallness of FSI effects has been put
forward by Bjorken \cite{Bjorken} based on the color-transparency
argument \cite{BL}. The renormalization-group (RG) analysis of soft gluon
exchanges among initial- and final-state mesons \cite{LT} has also
indicated that FSI effects are not important in two-body $B$ meson decays.
These discussions have led us to ignore FSI effects in the PQCD formalism.
For example, the charge exchange in the rescattering
$B^{+} \to K^{+}\pi^{0}\to K^{0}\pi^{+}$, regarded as occuring through
short-distance quark-pair annihilation, is of higher order \cite{KLS}.

As stated in the Introduction, the neglect of FSI effects requires
experimental justification. For this purpose, we propose to investigate
the $B\to KK$ decays, which are more sensitive to FSI effects compared 
with the $B\to K\pi$ and $\pi\pi$ decays. Similar proposals have been
presented in the literature \cite{He,GR} within the framework of $SU(3)$
symmetry. We make our argument explicit by means of the general
expression for the $B\to \pi\pi$, $K\pi$ and $KK$ decay amplitudes,
\begin{equation}
{\cal A}=V_{u}T+V_{u}P_{u}+V_{c}P_{c}+V_{t}P_{t}\;.
\label{amp}
\end{equation}
The factors $V_{q}=V_{qd}V_{qb}^{*}$, $q=u$, $c$, and $t$, are the
products of the Cabibbo-Kobayashi-Maskawa (CKM) matrix elements, $T$
denotes the tree amplitude, and $P_q$ denote the penguin amplitudes
arising from internal $q$-quark contributions. FSI effects have been
included in the amplitudes $T$ and $P_{u,c,t}$.

Using the unitarity relation $V_{c}=-V_{u}-V_{t}$, Eq.~(\ref{amp}) is
rewritten as
\begin{eqnarray}
{\cal A}&=&V_{u}\left(T+P_{u}-P_{c}\right)
+V_{t}\left(P_{t}-P_{c}\right)\;,
\nonumber\\
&=&V_{u}\left(T+P_{u}-P_{c}\right)\left[1+\frac{V_t}{V_u}
\frac{P_{t}-P_{c}}{T+P_{u}-P_{c}}\right]\;,
\nonumber\\
&\equiv&V_{u}\left(T+P_{u}-P_{c}\right)\left[1+\frac{V_t}{V_u}
{\cal R}_{\pi\pi(KK)}e^{i\delta_{\pi\pi(KK)}}\right]\;,
\label{fpp}
\end{eqnarray}
or
\begin{eqnarray}
{\cal A}&=&V_{t}\left(P_{t}-P_{c}\right) \left[ 1+\frac{V_{u}}{V_{t}}
\frac{T+P_{u}-P_{c}}{P_{t}-P_{c}} \right]\;,
\nonumber \\
&\equiv&V_{t}\left(P_{t}-P_{c}\right) \left[ 1+\frac{V_{u}}{V_{t}}
{\cal R}_{K\pi}e^{i\delta_{K\pi}}\right]\;,
\label{amp1}
\end{eqnarray}
where ${\cal R}$ are the ratios of different amplitudes and $\delta$ the
CP-conserving strong phases.

Without FSI, the various amplitudes $T$ and $P_{u,c,t}$, namely, the
ratios ${\cal R}$ and the strong phases $\delta$ are calculable in
PQCD. If FSI effects are important, they may change branching ratios
or induce CP asymmetries of two-body $B$ meson decays by varying $R$ and
$\delta$. For the $B\to \pi\pi$ decays, the ratio $V_t/V_u$ is of order
unity, but $R_{\pi\pi}$ is small because of the large Wilson coefficients
$a_1=C_2+C_1/N_c$ in $T$, $N_c$ being number of colors. The exception is
the $B_d^0\to \pi^0\pi^0$ mode, whose tree amplitude is proportional to
the small Wilson coefficient $a_2=C_1+C_2/N_c$. The ratio $R_{K\pi}$ may
be large, but its coefficient $V_u/V_t\sim R_{b}\lambda ^{2}$, $R_b$ and
$\lambda$ being the Wolfenstein parameters defined in Sec.~IV, is small.
Therefore, FSI effects in the $B\to \pi\pi$ and $K\pi$ decays are
suppressed by $1/a_1$ and $V_u/V_t$, respectively. On the other hand,
the $B\to \pi\pi$ and $K\pi$ decays have branching ratios of order
$10^{-5}$, which are larger than those of the $B\to KK$ decays (of order
$10^{-6}$ as calculated in Sec.~V). It has been also predicted in PQCD
that the CP asymmetries in the $B\to\pi\pi$ and $K\pi$ decays are large:
$30-40\%$ in the former \cite{LUY,KL} and $10-15\%$ in the latter
\cite{KLS}. These large values render FSI effects relatively mild.

For the $B\to KK$ decays, $T$ arises only from small nonfactorizable
annihilation diagrams for the $B^\pm\to K^\pm K^0$ and
$B_d^0\to K^\pm K^\mp$ modes, and vanishes for the
$B_d^0\to K^0 {\bar K}^0$ modes. Furthermore, $V_t/V_u$ is of order
unity. Hence, there is no suppression from the Wilson coefficients and
from the CKM matrix elements, and FSI effects will be more significant.
In the PQCD approach ${\cal R}_{KK}$ is close to unity, corresponding to
the branching ratios of order $10^{-6}$ for
$B^{\pm}\to K^\pm K^0$ and $B_d^0\to K^0 {\bar K}^0$, and $10^{-8}$ for
$B_d^{0}\to K^{\pm}K^{\mp}$. CP asymmetry vanishes in the
$B_d^{0}\to K^0{\bar K}^0$ modes, because only the penguin operators 
contribute at leading order. However, if FSI contributes, the above
results will be changed dramatically. For example, FSI effects could
induce large $T$ and $P_{u,c}$ via the rescattering of intermediate
states $DD$ and $\pi\pi$ produced from the tree operators, and $P_{t}$
via the rescattering of intermediate states $KK$ produced from the penguin
operators. When ${\cal R}_{KK}$ deviates from unity through rescattering
processes, the branching ratios and CP asymmetries of the $B\to KK$
decays could be enhanced.

We show how FSI effects modify amplitudes of various topologies in the
$B\to KK$ decays in Table I. For more allowed intermediate states, refer
to \cite{GR}. It is obvious that the rescattering processes
$DD(\pi\pi)\to KK$ may be important due to the large $B\to DD(\pi\pi)$
branching ratios. For example, $B(B_d^0\to \pi^\pm\pi^\pm)$ is of order
$10^{-5}$. It is then possible that FSI effects could be significant
enough to increase $B(B_d^0\to K^\pm K^\mp)$ from order $10^{-8}$ to
above $10^{-7}$. For a similar reason, the rescatering
processes could induce large $P_{u,c}$ with the CKM matrix elements
$V_{u,c}$, which, as interfered with the penguin contributions,
result in sizable CP asymmetry in the $B_d^0\to K^0 {\bar K}^0$ modes.
Hence, large CP asymmetry observed in the $B_d^0\to K^0 {\bar K}^0$ modes
and large deviation of the observed $B_d^0\to K^\pm K^\mp$
branching ratios from the PQCD predictions will indicate strong FSI
effects.

\section{NONFACTORIZABLE AND ANNIHILATION CONTRIBUTIONS}

PQCD factorization theorem for exclusive nonleptonic $B$ meson decays
has been briefly reviewed in \cite{KLS}. In this section we simply sketch
the idea of PQCD factorization theorem, concentrating on its application
to nonfactorizable and annihilation amplitudes in the $B\to KK$ decays.

In perturbation theory nonperturbative dynamics is reflected by infrared
divergences in radiative corrections. These infrared divergences can be
separated and absorbed into a $B$ meson wave function or a kaon wave
function order by order \cite{LY1}. A formal definition of the meson wave
functions as matrix elements of nonlocal operators can be constructed,
which, if evaluated perturbatively, reproduces the infrared divergences.
Certainly, one can not derive a wave function in perturbation theory, but
parametrizes it as a parton model, which describes how a parton (valence
quark, if a leading-twist wave function is referred) shares meson
momentum. The meson wave functions, characterized by the QCD scale
$\Lambda_{\rm QCD}$, must be determined by nonperturbative means, such as
lattice gauge theory and QCD sum rules, or extracted from experimental
data. In the application below small parton transverse momenta $k_T$ are
included, and the characteristic scale is replaced by $1/b$ with $b$
being a variable conjugate to $k_T$.

After absorbing infrared divergences into the meson wave functions, the
remaining part of radiative corrections is infrared finite. This part
can be evaluated perturbatively as a hard amplitude with six on-shell
external quarks, four of which correspond to the four-fermion operators
and two of which are the spectator quarks of the $B$ or $K$ mesons.
Note that the $b$ quark carries various momenta, whose distribution is
described by the $B$ meson wave function introduced above. The six-quark
amplitude contains all possible Feynman diagrams, which include both
factorizable and nonfactorizable tree, penguin and annihilation
contributions. A factorizable diagram involves hard gluon exchanges among
valence quarks of the $B$ meson or of a kaon. A nonfactorizable diagram
involves hard gluon exchanges bwteeen the valence quarks of different
mesons. That is,  the PQCD formalism does not rely on FA.

The hard amplitude is characterized by the virtuality $t$ of involved
internal particles, which is of order $M_B$, and by the $W$ boson mass
$M_W$. The hard scale $t$ reflects the specific dynamics of a decay mode,
while $M_W$ serves the scale, at which the matching conditions of the
effective weak Hamiltonian to the full Hamiltonian are defined. The
study of the pion form factor has indicated that the choice of $t$ as
the maximum of internal particle virtualities
minimizes next-to-leading-order corrections to hard amplitudes \cite{MN}.
Large logarithmic corrections are organized by RG methods. The results
consist of the evolution from $M_W$ down to $t$ described by the
Wilson coefficients, the evolution from $t$ to $1/b$, and a Sudakov
factor. The two evolutions are governed by different anomalous dimensions,
since loop corrections associated with spectator quarks contribute, when
the energy scale runs to below $t$. The Sudakov factor suppresses the
long-distance contributions from the large $b$ region, and vanishes as
$b=1/\Lambda_{\rm QCD}$. This suppression guarantees the applicability of
PQCD to exclusive decays around the energy scale of the $B$ meson mass
\cite{LY1}.

A salient feature of PQCD factorization theorem is the universality of
nonperturbative wave functions. Because of universality, meson wave
functions extracted from some decay modes can be employed to make
predictions for other modes. We have determined the $B$ and $K$ meson wave
functions from the experimental data of the $B\to K\pi$ and $\pi\pi$
decays \cite{KLS}, and the unitarity angle $\phi_3=90^o$ from the CLEO
data of the ratio \cite{CLEO3},
\begin{eqnarray}
R=\frac{B(B_d^0\to K^\pm\pi^\mp)}{B(B^\pm\to K^0\pi^\pm)}=0.95\pm 0.30\;,
\label{tr}
\end{eqnarray}
where $B(B_d^0\to K^\pm\pi^\mp)$ represents the CP average of the branching
ratios $B(B_d^0\to K^+\pi^-)$ and $B({\bar B}_d^0\to K^-\pi^+)$.
It has been emphasized that the $B\to\pi\pi$ data can be explained using
the same angle $\phi_3=90^o$ in PQCD, contrary to the conclusion in
\cite{WS,HY}, where an angle larger than $100^o$ must be adopted. In this
work we shall predict the branching ratios and CP asymmetries of the
$B\to KK$ decays in the PQCD formalism employing the above meson wave
functions and the unitarity angle.

Factorizable annihilation contributions correspond to the time-like kaon
form factor. It is known that annihilation contributions from the
$O_{1-4}$ operators with the $(V-A)(V-A)$ structure vanish because of
helicity suppression. However, those from the $O_{5,6}$ operators with the
$(V-A)(V+A)$ structure bypass helicity suppression, and turn out to
be comparible with penguin contributions \cite{KLS}. Without FSI in PQCD,
strong phases arise from non-pinched singularities of quark and gluon
propagators in annihilation and nonfactorizable diagrams. Especially,
annihilation amplitudes are the main source of strong phases
\cite{KLS}. In the FA and Beneke-Buchalla-Neubert-Sachrajda (BBNS)
\cite{BBNS,Du} approaches, where annihilation diagrams are not taken into
account, strong phases come from the Bander-Silverman-Soni (BSS)
mechanism \cite{BSS} and from the extraction of the scale dependence
from hadronic matrix elements \cite{Ali}. As shown in \cite{KL}, these
sources are in fact next-to-leading-order. As a consequence, CP
asymmetries predicted in FA and BBNS are smaller than those predicted in
PQCD. Nonfactorizable amplitudes have been also considered in the BBNS
approach, which are, however, treated in a different way. For example,
they are real because of some approximation in \cite{BBNS}, but complex
in PQCD \cite{WYL}. Generally speaking, nonfactorizable contributions
are less important compared to factorizable ones except in the cases
where factorizable contributions are proportional to the small
Wilson coefficients $a_2$ or vanish.

As stated before, the $B^\pm\to K^\pm K^0$ decays involve both
annihilation amplitudes from $O_{5,6}$ and nonfactorizable annihilation
amplitudes from $O_{1,2}$. Their interference then leads to substantial CP
asymmetry in PQCD. The $B_d^0\to K^\pm K^\mp$ decays involve only
nonfactorizable annihilation amplitudes from tree and penguin operators,
such that their branching ratios can not be estimated, or are vanishingly
small in the FA and BBNS approaches. These quantities mark the
essential differences among FA, BBNS and PQCD. The comparision of our
predictions for the CP asymmetry in the $B^\pm\to K^\pm K^0$ decays and
for the $B_d^0\to K^\pm K^\mp$ branching ratios with future data will
justify our evaluation of annihilation and nonfactorizable contributions,
and distinguish the FA, BBNS and PQCD approaches.

\section{FACTORIZATION FORMULAS}

We present the factorization formulas of the $B\to KK$ decays in this
section. The effective Hamiltonian for the flavor-changing $b\to d$
transition is given by \cite{REVIEW}
\begin{equation}
H_{\rm eff}={G_F\over\sqrt{2}}
\sum_{q=u,c}V_q\left[C_1(\mu)O_1^{(q)}(\mu)+C_2(\mu)O_2^{(q)}(\mu)+
\sum_{i=3}^{10}C_i(\mu)O_i(\mu)\right]\;,
\label{hbk}
\end{equation}
with the CKM matrix elements $V_q=V^*_{qd}V_{qb}$ and the operators
\begin{eqnarray}
& &O_1^{(q)} = (\bar{d}_iq_j)_{V-A}(\bar{q}_jb_i)_{V-A}\;,\;\;\;\;\;\;\;\;
O_2^{(q)} = (\bar{d}_iq_i)_{V-A}(\bar{q}_jb_j)_{V-A}\;,
\nonumber \\
& &O_3 =(\bar{d}_ib_i)_{V-A}\sum_{q}(\bar{q}_jq_j)_{V-A}\;,\;\;\;\;
O_4 =(\bar{d}_ib_j)_{V-A}\sum_{q}(\bar{q}_jq_i)_{V-A}\;,
\nonumber \\
& &O_5 =(\bar{d}_ib_i)_{V-A}\sum_{q}(\bar{q}_jq_j)_{V+A}\;,\;\;\;\;
O_6 =(\bar{d}_ib_j)_{V-A}\sum_{q}(\bar{q}_jq_i)_{V+A}\;,
\nonumber \\
& &O_7 =\frac{3}{2}(\bar{d}_ib_i)_{V-A}\sum_{q}e_q(\bar{q}_jq_j)_{V+A}\;,
\;\;
O_8 =\frac{3}{2}(\bar{d}_ib_j)_{V-A}\sum_{q}e_q(\bar{q}_jq_i)_{V+A}\;,
\nonumber \\
& &O_9 =\frac{3}{2}(\bar{d}_ib_i)_{V-A}\sum_{q}e_q(\bar{q}_jq_j)_{V-A}\;,
\;\;
O_{10} =\frac{3}{2}(\bar{d}_ib_j)_{V-A}\sum_{q}e_q(\bar{q}_jq_i)_{V-A}\;,
\end{eqnarray}
$i$ and $j$ being the color indices. Using the unitarity condition, the
CKM matrix elements for the penguin operators $O_3$-$O_{10}$ can also
be expressed as $V_u+V_c=-V_t$. The unitarity angle $\phi_3$ is defined
via
\begin{equation}
V_{ub}=|V_{ub}|\exp(-i\phi_3)\;.
\end{equation}
Adopting the Wolfenstein parametrization for the
CKM matrix upto ${\cal O}(\lambda^{3})$,
\begin{eqnarray}
\left(\matrix{V_{ud} & V_{us} & V_{ub} \cr
              V_{cd} & V_{cs} & V_{cb} \cr
              V_{td} & V_{ts} & V_{tb} \cr}\right)
=\left(\matrix{ 1 - { \lambda^2 \over 2 } & \lambda &
A \lambda^3(\rho - i \eta)\cr
- \lambda & 1 - { \lambda^2 \over 2 } & A \lambda^2\cr
A \lambda^3(1-\rho-i\eta) & -A \lambda^2 & 1 \cr}\right)\;,
\end{eqnarray}
we have the parameters \cite{PDG}
\begin{eqnarray}
\lambda &=& 0.2196 \pm 0.0023\;,
\nonumber \\
A &=& 0.819 \pm 0.035\;,
\nonumber \\
R_b &\equiv&\sqrt{{\rho}^2  + {\eta}^2}
= 0.41 \pm 0.07\;.
\end{eqnarray}

For the $B^\pm\to K^\pm K^0$ decays, the operators $O_{1,2}^{(u)}$
contribute via the annihilation topology, in which the fermion flow
forms two loops as shown in Fig.~1. $O_{1,2}^{(c)}$ do not contribute at
leading order of $\alpha_s$. $O_{3-10}$ contribute via the penguin
topology with the light quark $q=s$ and via the annihilation topology
with $q=u$, in which the fermion flow forms one loop. As evaluating hard
amplitudes, an additional minus sign should be associated with the
$O^{(u)}_{1,2}$ contributions, that contain two fermion loops.
$O_{1,3,5,7,9}$ give both factorizable and nonfactorizable
(color-suppressed) contributions, while $O_{2,4,6,8,10}$ give only
factorizable ones because of the color flow. The electroweak penguin
contributions from $O_{7-10}$ have been included in the same way
as those from the QCD penguins $O_{3-6}$. Obviously, the electroweak
penguins are less important because of the small electromagnetic
coupling.

The diagrams for the $B_d^0\to K^\pm K^\mp$ decays are displayed in Fig.~2.
The operators $O_{1,2}^{(u)}$ contribute via the annihilation topology,
which contain one fermion loop. $O_{1,2}^{(c)}$ do not contribute at
leading order of $\alpha_s$. $O_{3-10}$ contribute via the annihilation
topology with the light quark $q=s$ or $u$, in which the fermion flow
forms two loops. In these modes $O_{2,4,6,8,10}$ give both factorizable
and nonfactorizable contributions, while $O_{1,3,5,7,9}$ give only
factorizable ones because of the color flow. Only the operators
$O_{3-10}$ contribute to the $B_d^0\to K^0 {\bar K}^0$ modes via the
penguin topology with the light quark $q=s$ and via the annihilation
topology with the light quark $q=s$ or $d$. The penguin contributions
contain one fermion loop. The $q=s$ annihilation amplitudes involve two
fermion loops, while the $q=d$ annihilation amplitudes contain both cases
of one fermion loop and of two fermion loops as shown in Fig.~3.

The $B$ meson momentum in light-cone coordinates is chosen as
$P_1=(M_B/\sqrt{2})(1,1,{\bf 0}_T)$. Momenta of the two kaons are chosen
as $P_2=(M_B/\sqrt{2})(1,0,{\bf 0}_T)$ and $P_3=P_1-P_2$. We shall drop
the contributions of order $(M_K/M_B)^2\sim 5\%$, $M_K$ being the kaon
mass. The $B$ meson is at rest with the above parametrization of momenta.
We define the momenta of light valence quark in the $B$ meson as $k_1$,
where $k_1$ has a plus component $k_1^+$, giving the momentum fraction
$x_1=k_1^+/P_1^+$, and small transverse components ${\bf k}_{1T}$. The
two light valence quarks in the kaon involved in the $B\to K$ transition
form factor carry the longitudinal momenta $x_2P_2$ and $(1-x_2)P_2$, and
small transverse momenta ${\bf k}_{2T}$ and $-{\bf k}_{2T}$, respectively.
The two light valence quarks in the other kaon carry the longitudinal
momenta $x_3P_3$ and $(1-x_3)P_3$, and small transverse momenta
${\bf k}_{3T}$ and $-{\bf k}_{3T}$, respectively.

The Sudakov resummations of large logarithmic corrections to the $B$
and $K$ meson wave functions lead to the exponentials $\exp(-S_B)$,
$\exp(-S_{K_2})$ and $\exp(-S_{K_3})$, respectively, with the exponents
\begin{eqnarray}
S_B(t)&=&s(x_1P_1^+,b_1)+2\int_{1/b_1}^{t}\frac{d{\bar \mu}}{\bar \mu}
\gamma(\alpha_s({\bar \mu}))\;,
\nonumber \\
S_{K_2}(t)&=&s(x_2P_2^+,b_2)+s((1-x_2)P_2^+,b_2)+2\int_{1/b_2}^{t}
\frac{d{\bar \mu}}{\bar \mu}\gamma(\alpha_s({\bar \mu}))\;,
\nonumber\\
S_{K_3}(t)&=&s(x_3P_3^-,b_3)+s((1-x_3)P_3^-,b_3)+2\int_{1/b_3}^{t}
\frac{d{\bar \mu}}{\bar \mu}\gamma(\alpha_s({\bar \mu}))\;.
\label{sbk}
\end{eqnarray}
The variable $b_1$, $b_2$, and $b_3$ conjugate to the parton transverse
momentum $k_{1T}$, $k_{2T}$, and $k_{3T}$ represents the transverse
extent of the $B$ and $K$ mesons, respectively.
The exponent $s$ is written as \cite{CS,BS,LS}
\begin{equation}
s(Q,b)=\int_{1/b}^{Q}\frac{d \mu}{\mu}\left[\ln\left(\frac{Q}{\mu}
\right)A(\alpha_s(\mu))+B(\alpha_s(\mu))\right]\;,
\label{fsl}
\end{equation}
where the anomalous dimensions $A$ to two loops and $B$ to one loop are
\begin{eqnarray}
A&=&C_F\frac{\alpha_s}{\pi}+\left[\frac{67}{9}-\frac{\pi^2}{3}
-\frac{10}{27}f+\frac{2}{3}\beta_0\ln\left(\frac{e^{\gamma_E}}{2}\right)
\right]\left(\frac{\alpha_s}{\pi}\right)^2\;,
\nonumber \\
B&=&\frac{2}{3}\frac{\alpha_s}{\pi}\ln\left(\frac{e^{2\gamma_E-1}}
{2}\right)\;,
\end{eqnarray}
with $C_F=4/3$ a color factor, $f=4$ the active flavor number, and
$\gamma_E$ the Euler constant.
The one-loop expression of the running coupling constant,
\begin{equation}
\alpha_s(\mu)=\frac{4\pi}{\beta_0\ln(\mu^2/\Lambda_{\rm QCD}^2)}\;,
\end{equation}
is substituted into Eq.~(\ref{fsl}) with the coefficient
$\beta_{0}=(33-2f)/3$.
The anomalous dimension $\gamma=-\alpha_s/\pi$ describes the RG evolution
from $t$ to $1/b$.

The decay rates of $B^\pm\to K^\pm K^0$ have the expressions
\begin{equation}
\Gamma=\frac{G_F^2M_B^3}{128\pi}|{\cal A}|^2\;.
\label{dr1}
\end{equation}
The decay amplitudes ${\cal A}^+$ and ${\cal A}^-$ corresponding to
$B^+\to K^+K^0$ and $B^-\to K^-K^0$, respectively, are written as
\begin{eqnarray}
{\cal A^+}&=&f_KV_t^*F^{P(s)}_{46}+V_t^*{\cal M}^{P(s)}_{46}
+f_BV_t^*F^{P(u)}_{a6}+V_t^*{\cal M}^{P(u)}_{a46}-V_u^*{\cal M}_{a1}\;,
\label{Map}\\
{\cal A^-}&=&f_KV_tF^{P(s)}_{46}+V_t{\cal M}^{P(s)}_{46}
+f_BV_tF^{P(u)}_{a6}+V_t{\cal M}^{P(u)}_{a46}-V_u{\cal M}_{a1}\;,
\label{Mam}
\end{eqnarray}
with the kaon decay constant $f_K$. The notation $F$ (${\cal M}$)
represents factorizable (nonfactorizable) contributions, where the indices
$a$ and $P(q)$ denote the annihilation and penguin topologies,
respectively, with the $q$ quark pair emitted from the electroweak
penguins, and the subscripts 1, 4 and 6 label the Wilson coefficients
appearing in the factorization formulas. The nonfactorizable amplitude
${\cal M}_{a1}$ are from the operators $O^{(u)}_{1,2}$.

The decay rates of $B_d^0\to K^\pm K^\mp$ have the similar expressions
with the amplitudes
\begin{eqnarray}
{\cal A}&=&V_t^*({\cal M}^{P(u)}_{a35}+{\cal M}^{P(s)}_{a35})
-V_u^*{\cal M}_{a2}\;,
\label{Mbp}\\
{\bar{\cal A}}&=&V_t({\cal M}^{P(u)}_{a35}+{\cal M}^{P(s)}_{a35})
-V_u{\cal M}_{a2}\;,
\label{Mbm}
\end{eqnarray}
for $B_d^0\to K^+K^-$ and ${\bar B}_d^0\to K^-K^+$, respectively.
The notations are similar to those in Eqs.~(\ref{Map}) and (\ref{Mam}).
The decay amplitudes for $B_d^0\to K^0{\bar K}^0$ and
${\bar B}_d^0\to K^0{\bar K}^0$ are written as
\begin{eqnarray}
{\cal A}'&=&f_KV_t^*F^{P(s)}_{46}+V_t^*{\cal M}^{P(s)}_{46}
+f_BV_t^*F^{P(d)}_{a6}
+V_t^*({\cal M}^{P(d)}_{a46}+{\cal M}^{P(d)}_{a35}
+{\cal M}^{P(s)}_{a35})\;,
\label{Mcp}\\
{\bar{\cal A}}'&=&f_KV_tF^{P(s)}_{46}+V_t{\cal M}^{P(s)}_{46}
+f_BV_tF^{P(d)}_{a6}
+V_t({\cal M}^{P(d)}_{a46}+{\cal M}^{P(d)}_{a35}
+{\cal M}^{P(s)}_{a35})\;,
\label{Mcm}
\end{eqnarray}
respectively.

The factorizable contributions are written as
\begin{eqnarray}
F^{P(s)}_{46}&=&F^{P(s)}_{4}+F^{P(s)}_{6}\;,
\nonumber \\
F^{P(s)}_{4}&=&16\pi C_FM_B^2\int_0^1 dx_1dx_3\int_0^{\infty}
b_1db_1b_3db_3\phi_B(x_1,b_1)
\nonumber \\
& &\times\{\left[(1+x_3)\phi_K(x_3)+r_K(1-2x_3)\phi'_K(x_3)
\right]E^{(s)}_{e4}(t^{(1)}_e)h_e(x_1,x_3,b_1,b_3)
\nonumber\\
& &+2r_K\phi'_K(x_3)E^{(s)}_{e4}(t^{(2)}_e)
h_e(x_3,x_1,b_3,b_1)\}\;,
\label{int4}\\
F^{P(s)}_{6}&=&32\pi C_FM_B^2\int_0^1 dx_1dx_3\int_0^{\infty}
b_1db_1b_3db_3\phi_B(x_1,b_1)
\nonumber \\
& &\times r_K\{\left[\phi_K(x_3)+r_K(2+x_3)\phi'_K(x_3)
\right]E^{(s)}_{e6}(t^{(1)}_e)h_e(x_1,x_3,b_1,b_3)
\nonumber\\
& &+\left[x_1\phi_K(x_3)+2r_K(1-x_1)\phi'_K(x_3)\right]
E^{(s)}_{e6}(t^{(2)}_e)h_e(x_3,x_1,b_3,b_1)\}\;,
\label{int6}\\
F^{P(q)}_{a6}&=&32\pi C_FM_B^2\int_0^1 dx_2dx_3\int_0^{\infty}
b_2db_2b_3db_3
\nonumber \\
& &\times r_K\{\left[x_3\phi_K(1-x_2)\phi'_K(1-x_3)
+2\phi'_K(1-x_2)\phi_K(1-x_3)\right]
E^{(q)}_{a6}(t^{(1)}_a)h_a(x_2,x_3,b_2,b_3)
\nonumber\\
& &+\left[2\phi_K(1-x_2)\phi'_K(1-x_3)
+x_2\phi'_K(1-x_2)\phi_K(1-x_3)\right]
E^{(q)}_{a6}(t^{(2)}_a)h_a(x_3,x_2,b_3,b_2)\}\;,
\nonumber\\
& &
\label{exc6}
\end{eqnarray}
for the light quarks $q=u$ and $d$. The evolution factors are given by
\begin{eqnarray}
E^{(s)}_{ei}(t)&=&\alpha_s(t)a^{(s)}_i(t)\exp[-S_B(t)-S_{K3}(t)]\;,
\\
E^{(q)}_{ai}(t)&=&\alpha_s(t)a^{(q)}_i(t)\exp[-S_{K2}(t)-S_{K3}(t)]\;.
\end{eqnarray}
Notice the arguments $1-x_2$ and $1-x_3$ of the kaon wave functions
$\phi_K$ and $\phi'_K$ in Eqs.~(\ref{exc4}) and (\ref{exc6}). The
explicit expressions of the kaon wave functions will be given in Sec.~V,
where $x$ represents the momentum fraction of the light $u$ or $d$ quark.
However, to render the annihilation contributions for $q=s$ and for $q=u$
or $d$ have the same hard parts, we have labelled the $s$ quark momentum
by $x$ in the latter case, and changed the arguments of the kaon wave
functions to $1-x$.

The factorizable annihilation contribution associated
with the Wilson coefficient $a^{(q)}_4$ from Fig.~1(c)
is identical to zero because of helicity suppression as indicated by
\begin{eqnarray}
F^{P(q)}_{a4}&=&16\pi C_FM_B^2\int_0^1 dx_2dx_3\int_0^{\infty}
b_2db_2b_3db_3
\nonumber \\
& &\times\{\left[-x_3\phi_K(1-x_2)\phi_K(1-x_3)
-2r_K^2(1+x_3)\phi'_K(1-x_2)\phi'_K(1-x_3)\right]
\nonumber\\
& &\hspace{1.0 cm}\times
E^{(q)}_{a4}(t^{(1)}_a)h_a(x_2,x_3,b_2,b_3)
\nonumber\\
& &+\left[x_2\phi_K(1-x_2)\phi_K(1-x_3)
+2r_K^2(1+x_2)\phi'_K(1-x_2)\phi'_K(1-x_3)\right]
\nonumber\\
& &\hspace{1.0cm}\times
E^{(q)}_{a4}(t^{(2)}_a)h_a(x_3,x_2,b_3,b_2)\}\;.
\label{exc4}
\end{eqnarray}
The helicity suppression does not apply to the annihilation contributions
associated with $a^{(q)}_6$, and the two terms in Eq.~(\ref{exc6})
are constructive. It is easy to confirm these observations by
interchaning the integration variables $x_2$ and $x_3$ in the second
terms of Eqs.~(\ref{exc6}) and (\ref{exc4}). The factorization formulas
for $F_{a1}$ from Fig.~1(a) and for $F_{a2}$ from Fig.~2(a), associated
with the Wilson coefficient $a_1(t_a)$ and $a_2(t_a)$, respectively,
are the same as $F^{P(q)}_{a4}$, {\it i.e.}, vanish. The expressions of
$F^{P(q)}_{a35}$ from Figs.~2(b), 2(c), 3(c), and 3(d), associated with
the Wilson coefficients $a^{(q)}_3(t_a)+a^{(q)}_5(t_a)$, are also
the same as $F^{P(q)}_{a4}$ and vanish.

The hard functions $h$'s in Eqs~(\ref{int4})-(\ref{exc6}), are given by
\begin{eqnarray}
h_e(x_1,x_3,b_1,b_3)&=&K_{0}\left(\sqrt{x_1x_3}M_Bb_1\right)
\nonumber \\
& &\times \left[\theta(b_1-b_3)K_0\left(\sqrt{x_3}M_B
b_1\right)I_0\left(\sqrt{x_3}M_Bb_3\right)\right.
\nonumber \\
& &\left.+\theta(b_3-b_1)K_0\left(\sqrt{x_3}M_Bb_3\right)
I_0\left(\sqrt{x_3}M_Bb_1\right)\right]\;,
\label{dh}\\
h_a(x_2,x_3,b_2,b_3)&=&\left(\frac{i\pi}{2}\right)^2
H_0^{(1)}\left(\sqrt{x_2x_3}M_Bb_2\right)
\nonumber \\
& &\times\left[\theta(b_2-b_3)
H_0^{(1)}\left(\sqrt{x_3}M_Bb_2\right)
J_0\left(\sqrt{x_3}M_Bb_3\right)\right.
\nonumber \\
& &\left.+\theta(b_3-b_2)H_0^{(1)}\left(\sqrt{x_3}M_Bb_3\right)
J_0\left(\sqrt{x_3}M_Bb_2\right)\right]\;.
\label{ah}
\end{eqnarray}
The derivation of $h$, from the Fourier transformation of the
lowest-order $H$, is similar to that for the $B\to D\pi$ decays
\cite{YL,WYL}.
The hard scales $t$ are chosen as the maxima of the virtualities of
internal particles involved in $b$ quark decay amplitudes, including
$1/b_i$:
\begin{eqnarray}
t^{(1)}_e&=&{\rm max}(\sqrt{x_3}M_B,1/b_1,1/b_3)\;,
\nonumber\\
t^{(2)}_e&=&{\rm max}(\sqrt{x_1}M_B,1/b_1,1/b_3)\;,
\nonumber\\
t^{(1)}_a&=&{\rm max}(\sqrt{x_3}M_B,1/b_2,1/b_3)\;,
\nonumber\\
t^{(2)}_a&=&{\rm max}(\sqrt{x_2}M_B,1/b_2,1/b_3)\;,
\label{et}
\end{eqnarray}
which decrease higher-order corrections. The Sudakov factor in
Eq.~(\ref{sbk}) suppresses long-distance contributions from the large $b$
({\it i.e.}, large $\alpha_s(t)$) region, and improves the applicability
of PQCD to $B$ meson decays.

For the nonfactorizable amplitudes, the factorization formulas involve
the kinematic variables of all the three mesons, and the Sudakov exponent
is given by $S=S_B+S_{K2}+S_{K3}$. The integration over $b_3$ can be
performed trivially, leading to $b_3=b_1$ or $b_3=b_2$. Their
expressions are
\begin{eqnarray}
{\cal M}^{P(s)}_{46}&=&{\cal M}^{P(s)}_{4}+{\cal M}^{P(s)}_{6}\;,
\nonumber \\
{\cal M}^{P(s)}_{4}&=&-32\pi C_F\sqrt{2N_c}M_B^2\int_0^1 [dx]
\int_0^{\infty}b_1 db_1 b_2 db_2\phi_B(x_1,b_1)\phi_K(x_2)
\nonumber \\
& &\times \{\left[(x_1-x_2)\phi_K(x_3)
+r_K x_3\phi'_K(x_3)\right]E^{(s)\prime}_{e4}(t^{(1)}_d)
h^{(1)}_d(x_1,x_2,x_3,b_1,b_2,b_1)
\nonumber \\
& &+\left[(1-x_1-x_2+x_3)\phi_K(x_3)-r_K x_3\phi'_K(x_3)\right]
E^{(s)\prime}_{e4}(t^{(2)}_d)h^{(2)}_d(x_1,x_2,x_3,b_1,b_2,b_1)\}\;,
\label{md3}\\
{\cal M}^{P(s)}_{6}&=&-32\pi C_F\sqrt{2N_c}M_B^2\int_0^1 [dx]
\int_0^{\infty}b_1 db_1 b_2 db_2\phi_B(x_1,b_1)\phi'_K(x_2)
\nonumber \\
& &\times r_K\{\left[(x_1-x_2)\phi_K(x_3)
+r_K (x_1-x_2-x_3)\phi'_K(x_3)\right]
\nonumber \\
& &\hspace{1.0 cm}\times
E^{(s)\prime}_{e6}(t^{(1)}_d)h^{(1)}_d(x_1,x_2,x_3,b_1,b_2,b_1)
\nonumber \\
& &+\left[(1-x_1-x_2)\phi_K(x_3)+r_K (1-x_1-x_2+x_3)
\phi'_K(x_3)\right]
\nonumber\\
& &\hspace{1.0 cm}\times E^{(s)\prime}_{e6}(t^{(2)}_d)
h^{(2)}_d(x_1,x_2,x_3,b_1,b_2,b_1)\}\;,
\label{md5}\\
{\cal M}^{P(q)}_{a46}&=&{\cal M}^{P(q)}_{a4}+{\cal M}^{P(q)}_{a6}\;,
\nonumber \\
{\cal M}^{P(q)}_{a4}&=&32\pi C_F\sqrt{2N_c}M_B^2\int_0^1 [dx]
\int_0^{\infty}b_1 db_1 b_2 db_2\phi_B(x_1,b_1)
\nonumber \\
& &\times \{\left[x_3\phi_K(1-x_2)\phi_K(1-x_3)
-r_K^2(x_1-x_2-x_3)\phi'_K(1-x_2)\phi'_K(1-x_3)\right]
\nonumber \\
& &\hspace{1.0cm}
\times E^{(q)\prime}_{a4}(t^{(1)}_f)h^{(1)}_f(x_1,x_2,x_3,b_1,b_2,b_2)
\nonumber \\
& &-\left[(x_1+x_2)\phi_K(1-x_2)\phi_K(1-x_3)
+r_K^2(2+x_1+x_2+x_3)\phi'_K(1-x_2)\phi'_K(1-x_3)\right]
\nonumber\\
& &\hspace{1.0cm}\times E^{(q)\prime}_{a4}(t^{(2)}_f)
h^{(2)}_f(x_1,x_2,x_3,b_1,b_2,b_2)\}\;,
\label{mf3}\\
{\cal M}^{P(q)}_{a6}&=&32\pi C_F\sqrt{2N_c}M_B^2\int_0^1 [dx]
\int_0^{\infty}b_1 db_1 b_2 db_2\phi_B(x_1,b_1)
\nonumber \\
& &\times \{\left[-r_K x_3\phi_K(1-x_2)\phi'_K(1-x_3)
-r_K(x_1-x_2)\phi'_K(1-x_2)\phi_K(1-x_3)\right]
\nonumber\\
& &\hspace{1.0 cm}\times
E^{(q)\prime}_{a6}(t^{(1)}_f)h^{(1)}_f(x_1,x_2,x_3,b_1,b_2,b_2)
\nonumber \\
& &-\left[r_K(2-x_3)\phi_K(1-x_2)\phi'_K(1-x_3)
-r_K(2-x_1-x_2)\phi'_K(1-x_2)\phi_K(1-x_3)\right]
\nonumber\\
& &\hspace{1.0cm}\times
E^{(q)\prime}_{a6}(t^{(2)}_f)h^{(2)}_f(x_1,x_2,x_3,b_1,b_2,b_2)\}\;,
\label{mf5}
\end{eqnarray}
with the definition $[dx]\equiv dx_1dx_2dx_3$ and $q=u$ and $d$.

The nonfactorizable amplitudes ${\cal M}^{P(q)}_{a35}$ are written as
\begin{eqnarray}
{\cal M}^{P(q)}_{a35}&=&{\cal M}^{P(q)}_{a3}
+{\cal M}^{P(q)}_{a5}\;,
\nonumber\\
{\cal M}^{P(q)}_{a5}&=&-32\pi C_{F}\sqrt{2N_c}M_B^2
\int_0^1[dx]\int_0^\infty b_{1}db_{1}b_{2}db_{2}\phi_B(x_1,b_1)
\nonumber \\
& &\times \{\left[(x_1-x_2)\phi_K(1-x_2)\phi_K(1-x_3)
+r_K^2(x_1-x_2-x_3)\phi'_K(1-x_2)\phi'_K(1-x_3)\right]
\nonumber \\
& &\hspace{1.0 cm}\times
E^{(q)\prime}_{a5}(t_f^{(1)})h_f^{(1)}(x_{i},b_{i})
\nonumber \\
& &+\left[x_3\phi_K(1-x_2)\phi_K(1-x_3)
+r_K^2(2+x_1+x_2+x_3)\phi'_K(1-x_2)\phi'_K(1-x_3)\right]
\nonumber \\
& &\hspace{1.0 cm}\times
E^{(q)\prime}_{a5}(t_f^{(2)})h_f^{(2)}(x_{i},b_{i})\}\;,
\label{ma5}
\end{eqnarray}
for $q=u$ and $d$. The evolution factors are given by
\begin{eqnarray}
E^{(s)\prime}_{ei}(t)&=&\alpha_s(t)a^{(s)\prime}_{i}(t)
\exp[-S(t)|_{b_3=b_1}]\;,
\\
E^{(q)\prime}_{ai}(t)&=&\alpha_s(t)a^{(q)\prime}_{i}(t)
\exp[-S(t)|_{b_3=b_2}]\;.
\end{eqnarray}
The expressions of ${\cal M}_{a1}$, ${\cal M}_{a2}$ and
${\cal M}^{P(q)}_{a3}$ are the same as ${\cal M}^{P(q)}_{a4}$ but with
the Wilson coefficients $a'_1(t_f)$, $a'_2(t_f)$, and $a'_3(t_f)$,
respectively. The expressions of ${\cal M}^{P(s)}_{a35}$ are the same as
${\cal M}^{P(q)}_{a35}$ but with the kaon wave functions
$\phi^{(\prime)}_K(1-x_i)$ replaced by $\phi^{(\prime)}_K(x_i)$, $i=2$
and 3, and with the $q$ quark replaced by the $s$ quark.
Notice the difference between the hard parts of ${\cal M}^{P(q)}_{a6}$
and ${\cal M}^{P(q)}_{a5}$, which are associated with the $O_{5-8}$
operators. The former (latter) corresponds to the fermion flow from the
$b$ quark to the $d$ quark in the kaon (the $\bar d$ quark in the $B$
meson), {\it i.e.}, the case with one fermion loop (two fermion loops).
That is, the nonfactorizable contributions
associated with the structure $(V-A)(V+A)$ distinguish these two cases,
while those associated with the structure $(V-A)(V-A)$ do not.

The functions $h^{(j)}$, $j=1$ and 2, appearing in
Eqs.~(\ref{md3})-(\ref{mf5}), are written as
\begin{eqnarray}
h^{(j)}_d&=& \left[\theta(b_1-b_2)K_0\left(DM_B
b_1\right)I_0\left(DM_Bb_2\right)\right. \nonumber \\
& &\quad \left.
+\theta(b_2-b_1)K_0\left(DM_B b_2\right)
I_0\left(DM_B b_1\right)\right]
\nonumber \\
&  & \times  K_{0}(D_{j}M_Bb_{2})\;,\;\;\;\;\;\;\;\;\;\;\;\;\;\;
\mbox{for $D^2_{j} \geq 0$}\;,
\nonumber  \\
&  & \times \frac{i\pi}{2} H_{0}^{(1)}(\sqrt{|D_{j}^2|}M_Bb_{2})\;,\;\;\;\;
 \mbox{for $D^2_{j} \leq 0$}\;,
\label{hjd}
\\
h^{(j)}_f&=& \frac{i\pi}{2}
\left[\theta(b_1-b_2)H_0^{(1)}\left(FM_B
b_1\right)J_0\left(FM_Bb_2\right)\right. \nonumber \\
& &\quad\left.
+\theta(b_2-b_1)H_0^{(1)}\left(FM_B b_2\right)
J_0\left(FM_B b_1\right)\right]
\nonumber \\
&  & \times
 K_{0}(F_{j}M_Bb_{1})\;,\;\;\;\;\;\;\;\;\;\;\;\;\;\;
\mbox{for $F^2_{j} \geq 0$}\;,
\nonumber\\
&  & \times \frac{i\pi}{2} H_{0}^{(1)}(\sqrt{|F_{j}^2|}M_Bb_{1})\;,
\;\;\;\; \mbox{for $F^2_{j} \leq 0$}\;,
\label{hjf}
\end{eqnarray}
with the variables
\begin{eqnarray}
D^{2}&=&x_{1}x_{3}\;,
\nonumber \\
D_{1}^{2}&=&F_1^2=(x_{1}-x_{2})x_{3}\;,
\nonumber \\
D_{2}^{2}&=&-(1-x_{1}-x_{2})x_{3}\;,
\nonumber \\
F^{2}&=&x_{2}x_{3}\;,
\nonumber \\
F_{2}^{2}&=&x_{1}+x_{2}+(1-x_{1}-x_{2})x_{3}\;.
\end{eqnarray}
For details of the derivation of $h^{(j)}$, refer to \cite{WYL}.
The hard scales $t^{(j)}$ are chosen as
\begin{eqnarray}
t^{(1)}_d&=&{\rm max}(DM_B,\sqrt{|D_1^2|}M_B,1/b_1,1/b_2)\;,
\nonumber \\
t^{(2)}_d&=&{\rm max}(DM_B,\sqrt{|D_2^2|}M_B,1/b_1,1/b_2)\;,
\nonumber \\
t^{(1)}_f&=&{\rm max}(FM_B,\sqrt{|F_1^2|}M_B,1/b_1,1/b_2)\;,
\nonumber \\
t^{(2)}_f&=&{\rm max}(FM_B,\sqrt{|F_2^2|}M_B,1/b_1,1/b_2)\;.
\end{eqnarray}

In the above expressions the Wilson coefficients are
defined by
\begin{eqnarray}
a_1&=&C_2+\frac{C_1}{N_c}\;,
\nonumber\\
a'_1&=&\frac{C_1}{N_c}\;,
\nonumber\\
a_2&=&C_1+\frac{C_2}{N_c}\;,
\nonumber\\
a'_2&=&\frac{C_2}{N_c}\;,
\nonumber\\
a^{(q)}_{3}&=&C_3+\frac{C_4}{N_c}+\frac{3}{2}e_q
\left(C_9+\frac{C_{10}}{N_c}\right)\;,
\nonumber\\
a^{(q)\prime}_{3}&=&\frac{1}{N_c}\left(C_4+\frac{3}{2}e_qC_{10}\right)\;,
\nonumber\\
a^{(q)}_{4}&=&C_4+\frac{C_3}{N_c}+\frac{3}{2}e_q
\left(C_{10}+\frac{C_9}{N_c}\right)\;,
\nonumber\\
a^{(q)\prime}_{4}&=&\frac{1}{N_c}\left(C_3+\frac{3}{2}e_q C_9\right)\;,
\nonumber\\
a^{(q)}_{5}&=&C_5+\frac{C_6}{N_c}+\frac{3}{2}e_q
\left(C_7+\frac{C_8}{N_c}\right)\;,
\nonumber\\
a^{(q)\prime}_{5}&=&\frac{1}{N_c}\left(C_6+\frac{3}{2}e_q C_8\right)\;,
\nonumber\\
a^{(q)}_{6}&=&C_6+\frac{C_5}{N_c}+\frac{3}{2}e_q
\left(C_{8}+\frac{C_7}{N_c}\right)\;,
\nonumber\\
a^{(q)\prime}_{6}&=&\frac{1}{N_c}\left(C_5+\frac{3}{2}e_q C_7\right)\;.
\label{wilc}
\end{eqnarray}
Both QCD and electroweak penguin contributions have been included
as shown in Eq.~(\ref{wilc}). It is expected that electroweak
penguin contributions are small, as concluded in \cite{AS}.

The pseudovector and pseudoscalar kaon wave functions $\phi_K$ and
$\phi'_K$ are defined by
\begin{eqnarray}
\phi_K(x)&=&\int\frac{dy^+}{2\pi}e^{-ixP_3^-y^+}\frac{1}{2}
\langle 0|{\bar u}(y^+)\gamma^-\gamma_5 s(0)|\pi\rangle\;,
\\
\frac{m_{0K}}{P_3^-}\phi'_K(x)&=&
\int\frac{dy^+}{2\pi}e^{-ixP_3^-y^+}\frac{1}{2}
\langle 0|{\bar u}(y^+)\gamma_5 s(0)|\pi\rangle\;,
\end{eqnarray}
respectively, satisfying the normalization
\begin{eqnarray}
\int_0^1 dx\phi_K(x)=\int_0^1 dx\phi'_K(x)=
\frac{f_K}{2\sqrt{2N_c}}\;.
\end{eqnarray}
The factor $r_K$,
\begin{eqnarray}
r_K=\frac{m_{0K}}{M_B}\;,\;\;\;\;m_{0K}=\frac{M_K^2}{m_s+m_d}\;,
\end{eqnarray}
with $m_s$ and $m_d$ being the masses of the $s$ and $d$ quarks,
respectively, is associated with the normalization of the pseudoscalar
wave function $\phi'_K$. Note that we have included the intrinsic $b$
dependence for the heavy meson wave function $\phi_B$ but not for the
kaon wave functions \cite{KLS}. As the transverse extent $b$ approaches
zero, the $B$ meson wave function $\phi_B(x,b)$ reduces to the standard
parton model $\phi_B(x)$, {\it i.e.}, $\phi_B(x)=\phi_B(x,b=0)$, which
satisfies the normalization
\begin{equation}
\int_0^1\phi_B(x)dx=\frac{f_B}{2\sqrt{2N_c}}\;.
\label{dco}
\end{equation}

\section{NUMERICAL ANALYSIS}

In the factorization formulas derived in Sec.~IV, the Wilson coefficients
evolve with the hard scale $t$ that depends on the internal kinematic
variables $x_i$ and $b_i$. Wilson coefficients at a scale $\mu < M_W$ are
related to the corresponding ones at $\mu = M_W$ through usual RG
equations. Since the typical scale $t$ of a hard amplitude is smaller
than the $b$ quark mass $m_b=4.8$ GeV, we further evolve the Wilson
coefficients from $\mu=m_b$ down to $\mu = t$. For the scale $t$ below
the $c$ quark mass $m_c = 1.5$ GeV, we still employ the evolution function
with $f = 4$, instead of with $f =3$, for simplicity, since the matching
at $m_c$ is less essential. Therefore, we set $f=4$ in the RG evolution
between $t$ and $1/b$ governed by the quark anomalous dimension $\gamma$.
The explicit expressions of $C_i(\mu)$ are referred to \cite{KLS}.

For the $B$ meson wave function, we adopt the model \cite{KLS}
\begin{eqnarray}
\phi_B(x,b)=N_Bx^2(1-x)^2
\exp\left[-\frac{1}{2}\left(\frac{xM_B}{\omega_B}\right)^2
-\frac{\omega_B^2 b^2}{2}\right]\;,
\label{os}
\end{eqnarray}
with the shape parameter $\omega_B=0.4$ GeV \cite{BW}. The normalization
constant $N_B=91.7835$ GeV is related to the decay constant $f_B=190$ MeV.
The kaon wave functions are chosen as
\begin{eqnarray}
\phi_{K}(x)&=&\frac{3}{\sqrt{2N_c}}f_{K}
x(1-x)[1+0.51(1-2x)+0.3(5(1-2x)^2-1)]\;,
\label{mn}\\
\phi'_K(x)&=&\frac{3}{\sqrt{2N_c}}f_{K}x(1-x)\;.
\end{eqnarray}
$\phi_K$ is derived from QCD sum rules \cite{PB2}, where the second term
$1-2x$, rendering $\phi_K$ a bit asymmetric, corresponds to $SU(3)$
symmetry breaking effect. The decay constant $f_K$ is set to 160 MeV
(in the convention $f_\pi=130$ MeV). The wave funcitons $\phi_B$ and
$\phi'_K$ were determined from the data of the $B\to K\pi$ decays
\cite{KLS}.

We employ $G_F=1.16639\times 10^{-5}$ GeV$^{-2}$, the Wolfenstein
parameters $\lambda=0.2196$, $A=0.819$, and $R_b=0.38$, the unitarity
angle $\phi_3= 90^o$, the masses $M_B=5.28$ GeV, $M_K=0.49$ GeV, and
$m_s=100$ MeV, which correspond to $m_{0K}=2.22$ GeV \cite{KLS}, and the
${\bar B}_d^0$ ($B^-$) meson lifetime $\tau_{B^0}=1.55$ ps
($\tau_{B^-}=1.65$ ps) \cite{PDG}. Our predictions for the branching
ratio of each mode are
\begin{eqnarray}
& &B(B^+\to K^+K^0)=1.47\times 10^{-6}\;,
\nonumber\\
& &B(B^-\to K^-K^0)=1.84\times 10^{-6}\;,
\nonumber\\
& &B(B_d^0\to K^+K^-)=3.27\times 10^{-8} \;,
\nonumber\\
& &B({\bar B}_d^0\to K^-K^+)=5.90\times 10^{-8}\;,
\nonumber\\
& &B(B_d^0\to K^0{\bar K}^0)=1.75\times 10^{-6} \;,
\nonumber\\
& &B({\bar B}_d^0\to K^0{\bar K}^0)=1.75\times 10^{-6}\;.
\end{eqnarray}
The above values are lower than those of the $B\to\pi\pi$ decays
\cite{LUY,KL}. Since the $B_d^0\to K^\pm K^\mp$ modes involved only
nonfactorizable annihilation amplitudes, their branching ratios are much
smaller than those of the $B^\pm\to K^\pm K^0$ and
$B_d^0\to K^0 {\bar K}^0$ modes. As explained in Sec.~II, a large
deviation of future experimental data from the predicted
$B_d^0\to K^\pm K^\mp$ branching ratios will imply the existence of
large FSI effects.

So far, CLEO gives only the upper bound of the $B\to KK$ decays
\cite{CLEO3}:
\begin{eqnarray}
& &B(B^\pm\to K^\pm K^0)< 5.1\times 10^{-6}\;,
\nonumber\\
& &B(B_d^0\to K^\pm K^\mp)< 2.0\times 10^{-6}\;.
\end{eqnarray}
We also quote the upper bound
\begin{eqnarray}
B(B_d^0\to K^0K^0)< 1.7\times 10^{-5} \;,
\end{eqnarray}
from \cite{PDG}. Obviously, our preictions are consistent with the above
data.

The CP asymmetries are defined by
\begin{equation}
A_{CP}=\frac{B({\bar B}\to KK)-B(B\to KK)}
{B({\bar B}\to KK)+B(B\to KK)}\;.
\end{equation}
Employing the above set of parameters and $\phi_3=90^o$, we predict
\begin{eqnarray}
& &A_{CP}(B^\pm\to K^\pm K^0)=0.11\;,
\nonumber \\
& &A_{CP}(B_d^0\to K^\pm K^\mp)=0.29\;,
\nonumber \\
& &A_{CP}(B_d^0\to K^0 K^0)=0\;.
\end{eqnarray}
Basically, the values are of the same order of those in the $B\to K\pi$
decays \cite{KLS}. The CP asymmetry in the $K^0 {\bar K}^0$ modes
vanishes, because they involve only penguin contributions. Measurements
of the CP asymmetry in the $B_d^\pm\to K^\pm K^0$ can justify the PQCD
evaluation of annihilation and nonfactorizable contributions to two-body
$B$ meson decays, and distinguish the FA, BBNS and PQCD approaches. The
significant CP asymmetry observed in the $B_d^0\to K^0 K^0$ will indicate
strong FSI effects.

The dependences of the $B\to KK$ branching ratios on the angle $\phi_3$
are displayed in Fig.~4. The branching ratios of the $K^\pm K^0$ modes
increase with $\phi_3$, while those of the $K^{\pm} K^{\mp}$ modes
decrease with $\phi_3$. The branching ratios of the $K^0 {\bar K}^0$
modes are insensitive to the variation of $\phi_3$. The variation with
$\phi_3$ is mainly a consequence of the inteference between the penguin
contributions and the nonfactorizable annihilation contributions
${\cal M}_a$ from the tree operators. Since ${\cal M}_{a1}$ in
Eqs.~(\ref{Map}) and (\ref{Mam}) and ${\cal M}_{a2}$ in
Eqs.~(\ref{Mbp}) and (\ref{Mbm}) contain the Wilson coefficients
$a'_1$ and $a'_2$, respectively, which are opposite in sign, the
behaviors of the branching ratios with $\phi_3$ in Figs.~4(a) and 4(b)
are different.

The dependences of the CP asymmetries on the angle $\phi_3$ are displayed
in Fig.~5. The CP asymmetry in the $K^0 {\bar K}^0$ modes
remains vanishing. The CP asymmetry in the $K^\pm K^\mp$ modes drop
suddenly from 70\% to zero near the high end of $\phi_3$. Since their
branching ratios and the denominator in the definition of $A_{CP}$ are
small, the variation with $\phi$ is amplified. Figures 4 and 5 can be
employed to determine the range of the angle $\phi_3$, when compared
with future data.

\section{CONCLUSION}

In this paper we have predicted the branching ratios and the CP
asymmetries of all the $B\to KK$ modes using PQCD factorization theorem.
The unitarity angle $\phi_3= 90^o$ and the universal $B$ and $K$ meson wave
functions extracted from the data of the $B\to K\pi$ and
$\pi\pi$ decays have been employed. The dependences of the branching
ratios and the CP asymmetries on the angle $\phi_3$ have been also
presented. These predictions can be confronted with future experimental
data. We believe that these modes can be observed in $B$ factories,
which have started their operation recently.

The $B\to KK$ decays are very important for understanding dynamics of
nonleptonic two-body $B$ meson decays, such as FSI, annihilation and
nonfactorizable effects. In the PQCD formalism FSI effects have been
assumed to be small. As explained in Sec.~II, the $B\to KK$ decays are
more sensitive to these effects compared to the $B\to K\pi$ and $\pi\pi$
decays. Hence, the comparision of our predictions, especially for the CP
asymmetry in the $B_d^0\to K^0 K^0$ decays and for the
$B_d^0\to K^\pm K^\mp$ branching ratios, with future data provides a
justification of the assumption. In PQCD the CP asymmetry of the
$B^\pm\to K^\pm K^0$ modes depends on annihilation amplitudes. It has
been argued that CP asymmetries in the $B\to KK$ decays are small in the
FA and BBNS approaches, where annihilation contributions have been
neglected. Therefore, experimental data of CP asymmetries will
distinguish the FA, BBNS and PQCD approaches. The $B_d^0\to K^\pm K^\mp$
modes involve only nonfactorizable annihilation amplitudes, such that
their branching ratios can not be estimated in FA and BBNS. Future data
of these modes can also verify the PQCD evaluation of the above
contributions.

\vskip 1.0cm

We thank X.G. He, Y.Y. Keum and A.I. Sanda for useful duscussion.
This work was supported in part by the National Center of Theoretical
Science, by the National Science Council of R.O.C. under the Grant No.
NSC-89-2112-M-006-004, and by the Grant-in Aid for Scientific
Exchange from the Ministry of Education, Science and Culture of Japan.

\vskip 1.0cm

\vskip 1.0cm

\begin{table}

\begin{tabular}{|l|l|l|l|l|}
\hline
modes & intermediate & affected & branching ratios for & data of \\
& states & topologies & intermediate states \cite{YK,PDG}
& branching ratios \\ \hline
$B^+\rightarrow K^+{\bar K}^0$ & ${\bar D}^+D^0$ & $P_c$ &
$<6.7\times 10^{-3}$ & $<5.1\times 10^{-6}$ \\ \hline
$B_d^0\rightarrow K^+K^-$ & $\pi^+\pi^-$ & $T,P_u$ & $
4.7_{-1.5}^{+1.8}\pm 0.6\times 10^{-6}$ & $<2.0\times 10^{-6}$ \\
\cline{2-4} & $\pi^0\pi^0$ & $T,P_u$ & $<9.3\times 10^{-6}$ &
\\ \cline{2-4} & $K^0{\bar K}^0$ & $P_t$ &
$<1.7\times 10^{-5}$ &  \\ \hline
$B_d^0\rightarrow K^0{\bar K}^0$ & $D^+D^-$ & $P_c$ &
$<5.9\times 10^{-3}$ & $<1.7\times 10^{-5}$
\\ \cline{2-4} & $\pi^+\pi^-$ & $P_u$ &
$4.7_{-1.5}^{+1.8}\pm 0.6\times 10^{-6}$ &
\\ \cline{2-4}
& $\pi^0\pi^0$ & $P_u$ & $<9.3\times 10^{-6}$ &  \\ \hline 
\end{tabular}

\caption{FSI effects in the $B\to KK$ decays.}

\end{table}

\vskip 2.0cm

{\bf \Large Figure Captions}
\vspace{10mm}

\begin{enumerate}
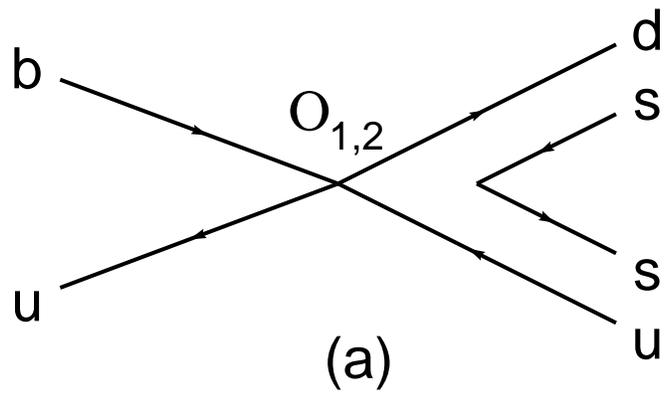
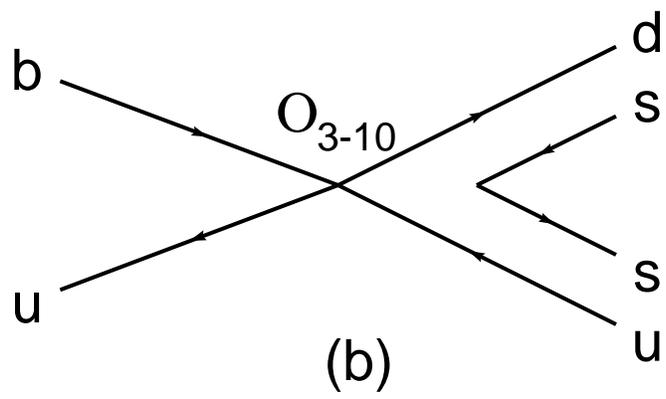
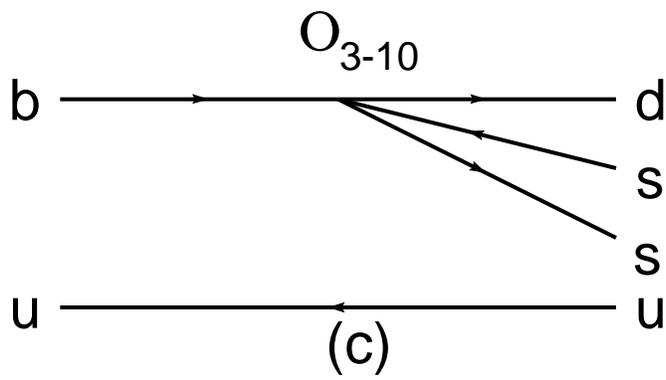
\item Fig. 1: Feynman diagrams for the $B^{\pm}\to K^\pm K^0$ decays.

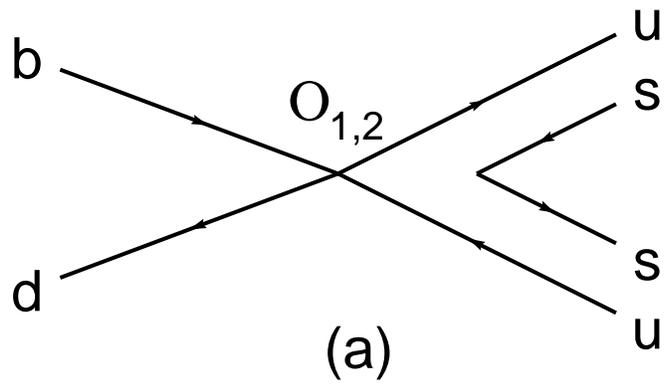
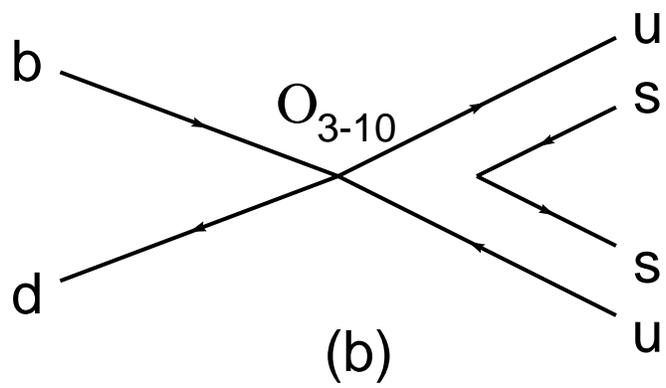
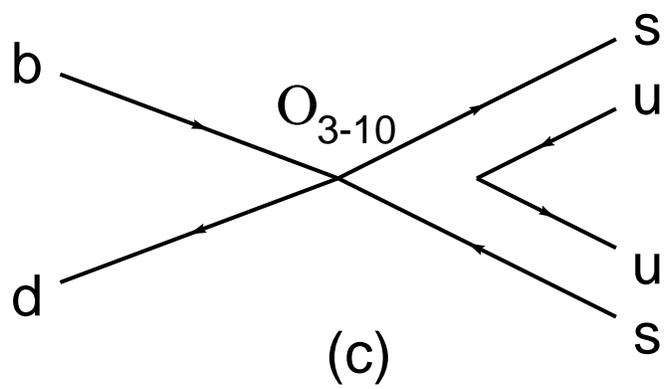
\item Fig. 2: Feynman diagrams for the $B_d^0\to K^\pm K^\mp$ decays.

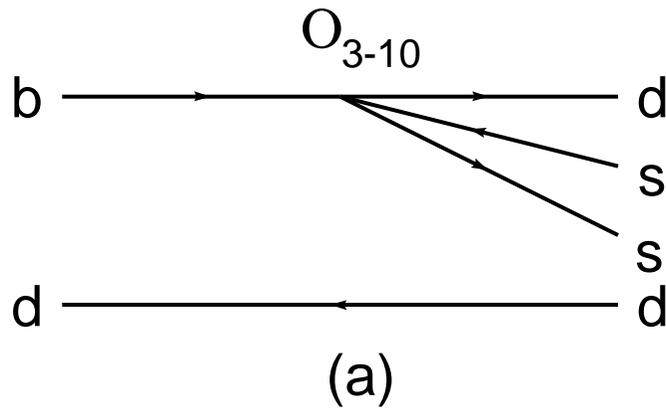
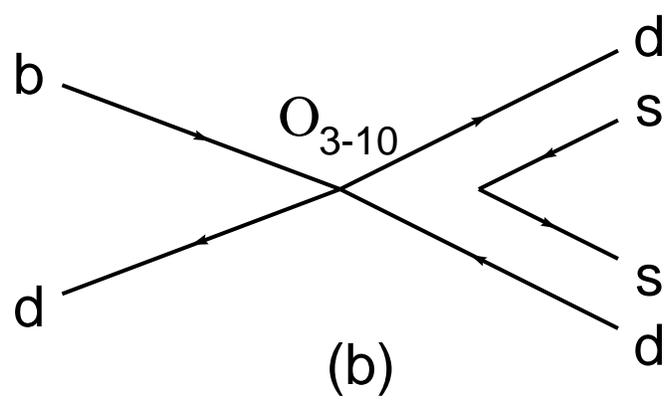
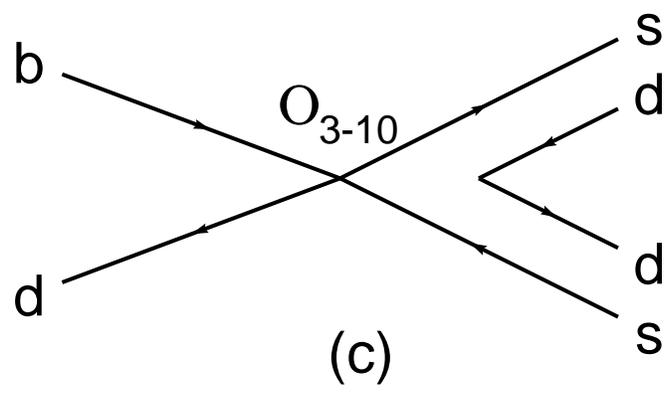
\item Fig. 3: Feynman diagrams for the $B_d^0\to K^0 {\bar K}^0$ decays.

\item Fig. 4: Dependences of the branching ratios on $\phi_3$ for
(a) the $B^{\pm}\to K^\pm K^0$ modes, (b) the $B_d^0\to K^\pm K^\mp$
modes and (c) the $B_d^0\to K^0 {\bar K}^0$ modes. The upper
(lower) lines correspond to the ${\bar B}$ ($B$) meson decays.

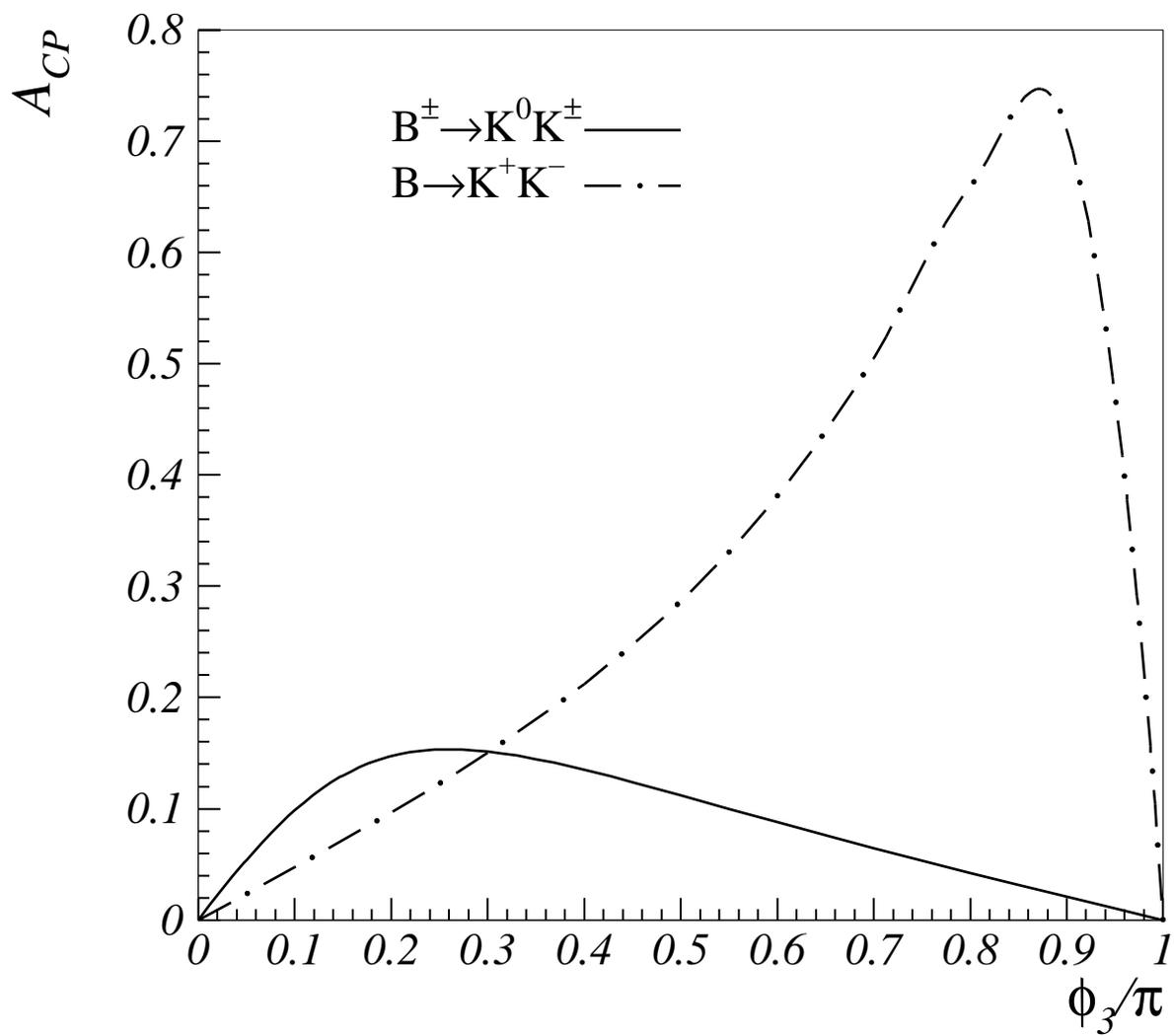
\item Fig. 5: Dependence of CP asymmetries on $\phi_3$ for
(a) the $B^{\pm}\to K^\pm K^0$ modes, (b) the $B_d^0\to K^\pm K^\mp$
modes and (c) the $B_d^0\to K^0 {\bar K}^0$ modes.

\end{enumerate}

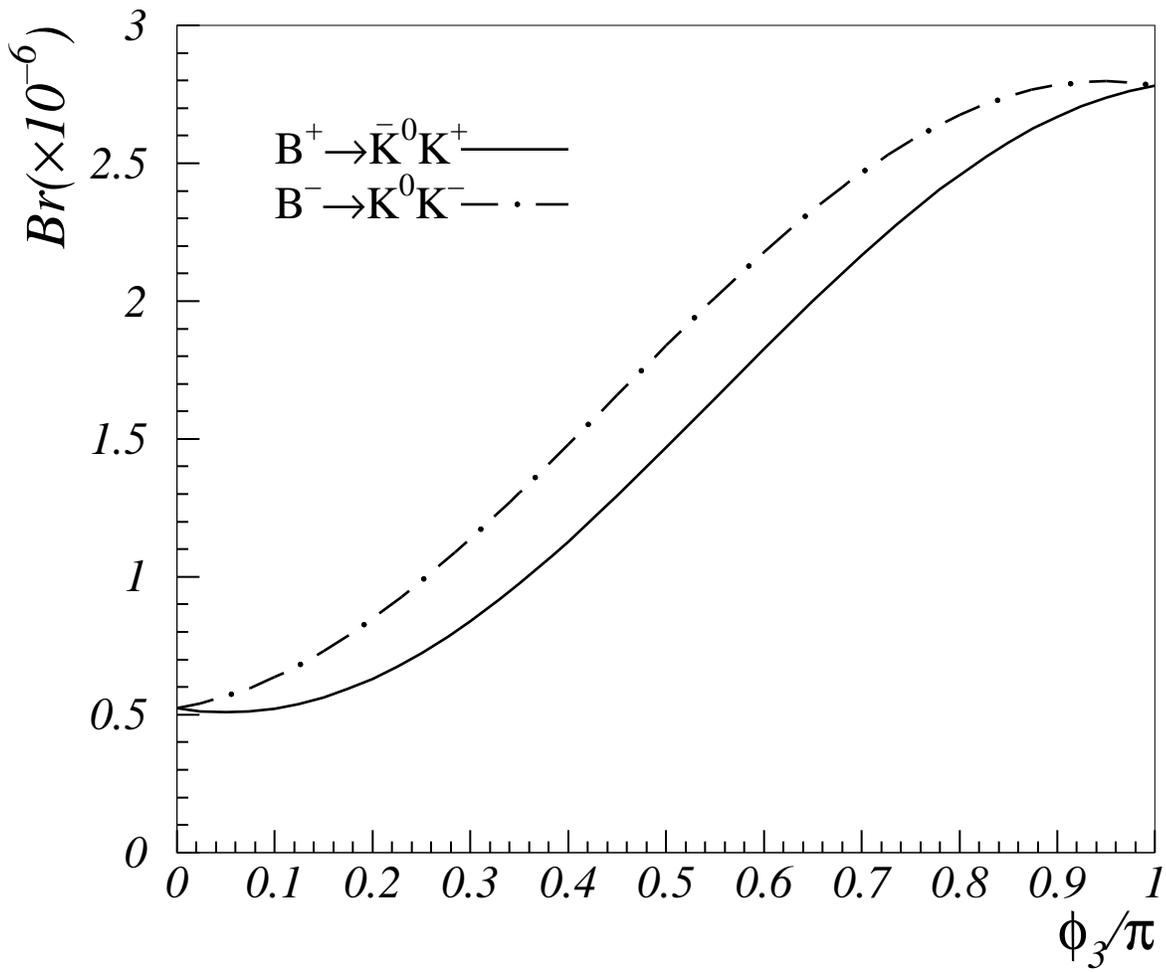
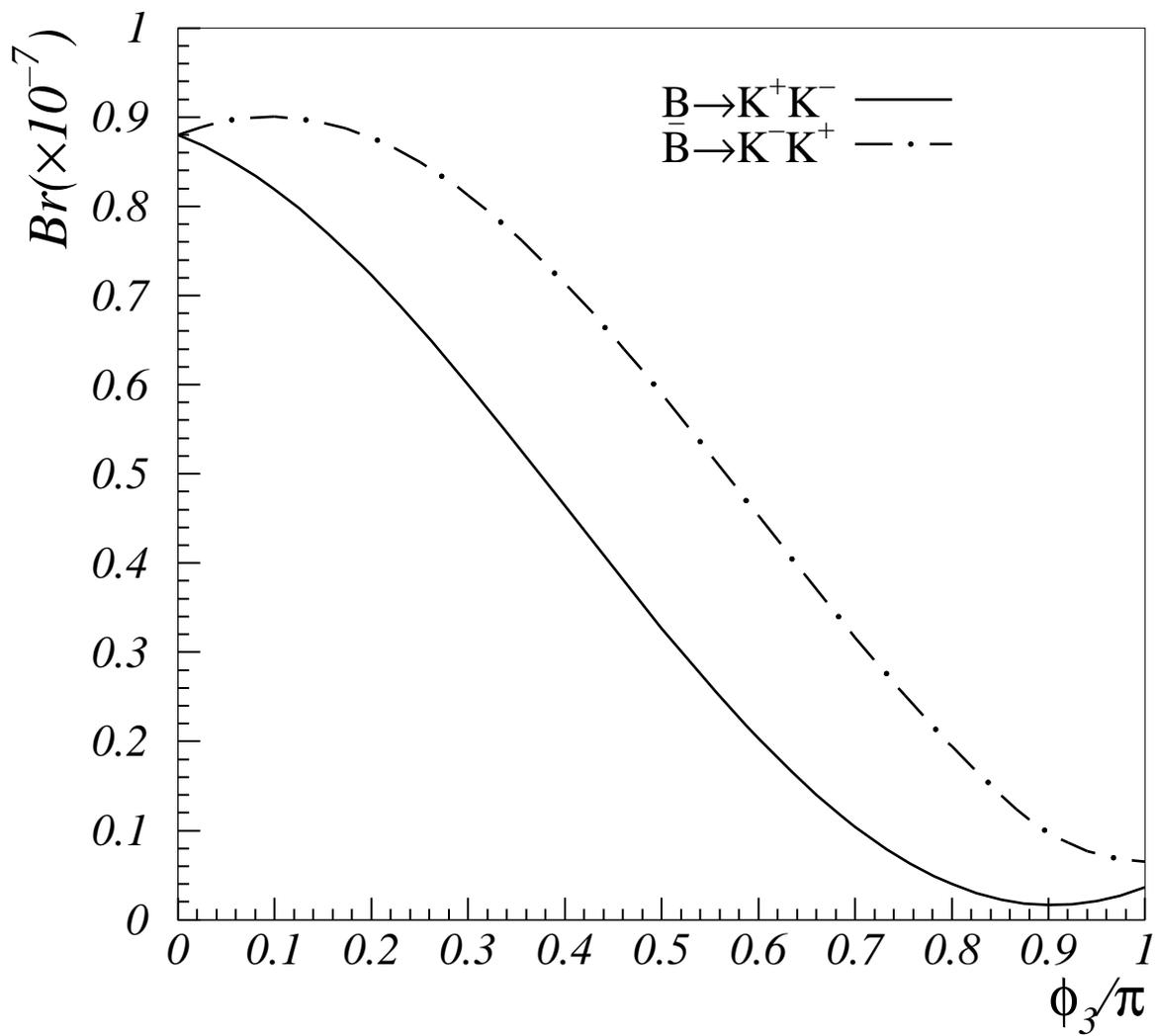
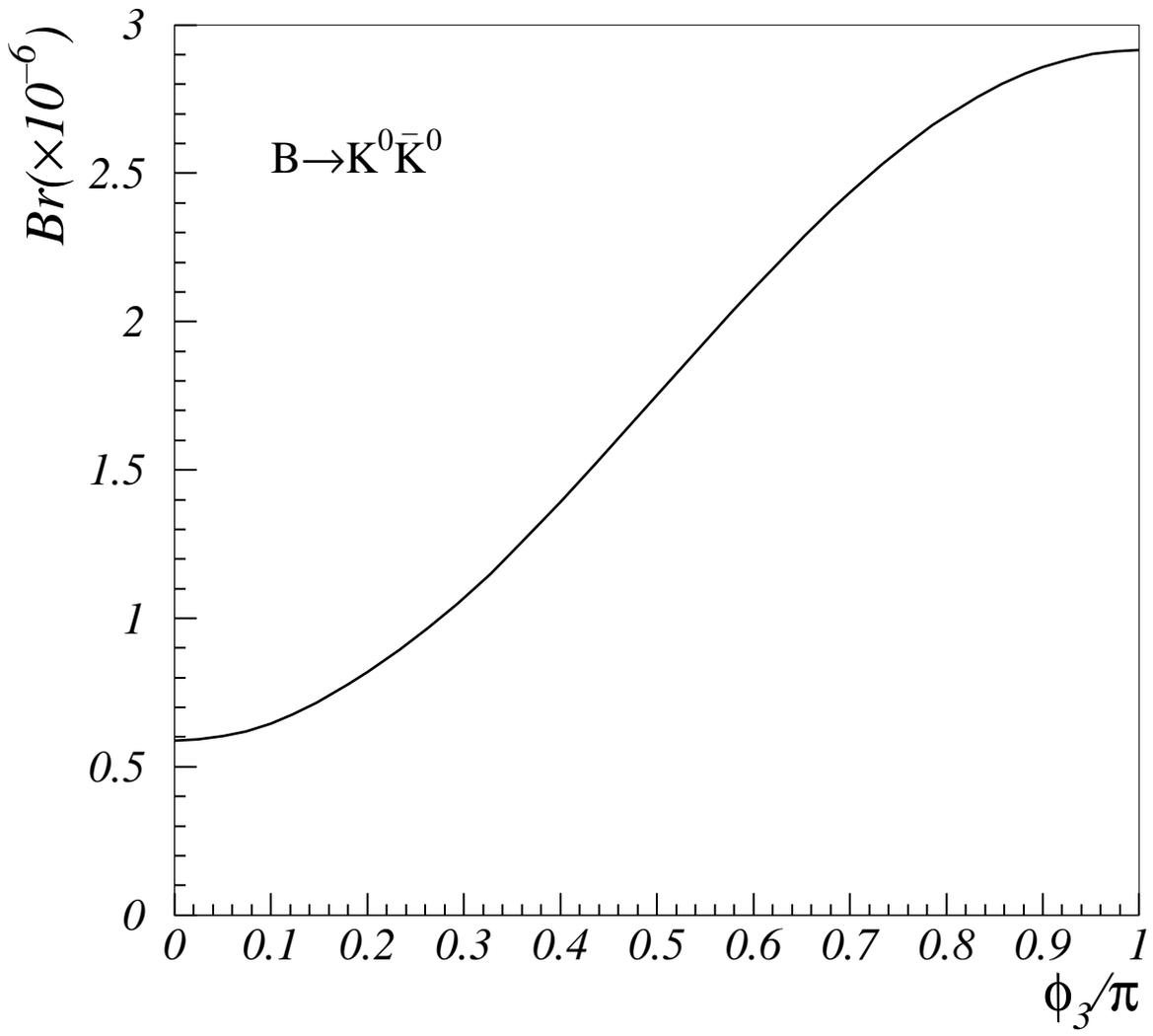
\end{document}